# Gamow-Teller transitions from $^{24}$Mg and its impact on the electron capture rates in the O + Ne + Mg cores of stars.

**Jameel-Un Nabi[*], Muneeb-Ur Rahman**

Faculty of Engineering Sciences, GIK Institute of Engineering Sciences and Technology, Topi 23640, Swabi, NWFP, Pakistan

**Abstract:**

Electron captures on nuclei play an important role in the collapse of stellar core in the stages leading to a type-II supernova. Recent observations of subluminous Type II-P supernovae (e.g. 2005cs, 2003gd, 1999br) were able to rekindle the interest in $8 - 10$ $M_\odot$ which develop O+Ne+Mg cores. We used the proton-neutron quasiparticle random phase approximation (pn-QRPA) theory to calculate the B(GT) strength for $^{24}$Mg $\rightarrow$ $^{24}$Na and its associated electron capture rates for incorporation in simulation calculations. The calculated rates, in this letter, have differences with the earlier reported shell model and Fuller, Fowler, Newman (hereafter $F^2N$) rates. We compared Gamow-Teller strength distribution functions and found fairly good agreement with experiment and shell model. However, the GT centroid and the total GT strength, which are useful in the calculation of electron capture rates in the core of massive pre-supernova stars, lead to the enhancement of our rate up to a factor of four compared to the shell model rates at high temperatures and densities.

PACS: 26.50.+x; 23.40.Bw; 23.40.-s; 21.60Jz

Keywords: Gamow-Teller strength; GT centroid, Electron capture rates, pn-QRPA, Core collapse

[*] Corresponding author
e-mail: jameel@giki.edu.pk
Fax: +92-938-271862

The spectacular visual display in which a star ends its life is known as supernova explosion where the core of the star collapses and produces energy of the order of $10^{46}$ J in the form of neutrinos. Electron capture on nuclei and on free protons plays an important role in the core collapse of a massive star. Electron capture and beta decay, during the final evolution of a massive star, are dominated by Fermi and Gamow-Teller transitions [1, 2]. The electron capture is very sensitive to the location and distribution of the $GT_+$ strength (in the $GT_+$ direction, a proton is changed into a neutron). The calculations of $F^2N$ [2] have shown that, for density exceeding $10^7$ g-cm$^{-3}$, electron capture transitions to the GT resonance are an important part of the rate.

In the late stages of the star evolution, energies of the electrons are high enough to induce transitions to the GT resonance. Experimentally the (p,n), (n,p), (d,$^2$He), and ($^3$He,t) reactions can be used to probe the GT transitions at higher excitation energy [3]. It was reported that the total $GT_+$ strength was quenched and distributed over many final states in daughter caused by the residual nucleon-nucleon correlation [4].

In addition to nuclear structure, GT transitions in nuclei directly affect the early phases of type-II supernova core collapse since the electron capture rates are partly determined by these GT transitions. The centroid of the GT distribution determines the effective energy of the electron capture and β-decay reactions. This along with the electron-Fermi energy determines which nuclei are able to capture electrons from, or β-decay onto the Fermi-sea at a given temperature and density and thus controls the rate at which the abundance of a particular nuclear species would change in the pre-supernova core.

The evolution of the stars in the mass range $8-10M_\odot$ develops central cores which are composed of $^{16}$O, $^{20}$Ne, and $^{24}$Mg. F$^2$N [2] compiled the experimental data and calculated electron capture rates for the nuclei in the mass range A = 21-60 for a wide grid of density and temperature. For the discrete transitions for which the experimental $ft$ values were not available $F^2N$ took $\log ft = 5.0$.

Later, Oda et al. [5] used the shell model wave functions of the sd-shell nuclei developed by Wildenthal [6] and calculated the electron capture rates which contribute to the collapse of O+Ne+Mg core. Oda et al. [5] pointed out three different series of electron capture in the O+Ne+Mg core of the $8-10M_\odot$ stars and placed them in the order of low threshold energy as $^{24}$Mg $\rightarrow$ $^{24}$Na $\rightarrow$ $^{24}$Ne, $^{20}$Ne $\rightarrow$ $^{20}$F $\rightarrow$ $^{20}$O, and $^{16}$O $\rightarrow$ $^{16}$N $\rightarrow$ $^{16}$C. They renounced the last series in their calculations because of no contribution to the initiation of the collapse of the O + Ne + Mg core of the $8-10M_\odot$ stars due to its high threshold energy.

We used the proton-neutron Quasi-particle Random Phase Approximation (pn-QRPA) theory to report here the electron capture rates for $^{24}$Mg nuclei in the central core of $8-10M_\odot$ stars. The idea is to provide the collapse simulators with a reliable alternate for a microscopic calculation of stellar weak rates which is one of the key input parameters to the simulation codes. We calculated the electron capture rates in the sd-shell for 178 nuclei from A = 17 to 40 using the Q values from the recent experimental mass compilation of Audi et al. [7]. The ratio of the axial-vector $\left(g_A\right)$ to the vector $\left(g_V\right)$ coupling constant was taken as -1.254. This value is consistent with reported value in the literature [8, 9]. Reliability of the weak rates is a key issue and of decisive importance for simulation codes. Compare to shell model calculations [5], the pn-QRPA gives similar accuracy in reproducing beta-decay rates in sd-shell nuclides [10, 11, 12]. There the authors compared the measured data (half lives and B(GT) strength) of thousands of nuclides with the pn-QRPA calculations and got fairly good comparison. (See especially Fig. 8 of [13]).

In this letter, we present Gamow-Teller strength distributions and associated electron capture rates, for $^{24}$Mg, using the pn-QRPA theory. The presence of $^{24}$Mg in the core is a result of the previous phase of carbon burning and its relevance is due to its lower electron capture threshold. To calculate the electron capture rates, we used the Hamiltonian

$$H^{QRPA} = H^{sp} + V^{pair} + V_{GT}^{ph} + V_{GT}^{pp}.$$ (2)

Here $H^{sp}$ is the single-particle Hamiltonian, $V^{pair}$ is the pairing force, $V_{GT}^{ph}$ is the particle-hole (ph) Gamow-Teller force, and $V_{GT}^{pp}$ is the particle-particle (pp) Gamow-Teller force. We calculated the single particle energies and wave functions in the Nilsson model [14], which takes into account the nuclear deformations. The proton-neutron residual interactions occur in particle-hole and particle-particle interaction forms. These interactions were characterized by two interaction constants $\chi$ and $\kappa$, respectively. In this work, we took the values of $\chi = 0.001$ MeV and $\kappa = 0.05$ MeV for $^{24}$Mg. For the detailed analysis of the Gamow-Teller strength parameters in pn-QRPA calculations we refer to [12, 13].

The decay rates from the $i$th state of the parent to the $j$th state of the daughter nucleus is given by

$$\lambda_{ij} = \ln 2 \frac{f_{ij}(T, \rho, E_f)}{(ft)_{ij}}.$$ (3)

Details of calculations of phase space integrals $f_{ij}$ and reduced transition probabilities can be found in [10, 15, 16]. We incorporated experimental data wherever available to further strengthen the reliability of our rates. The calculated excitation energies (along with their log $ft$ values) were replaced with experimental one when they were within 0.5MeV of each other. Missing measured states were inserted and inverse and mirror transitions were also taken into account. We did not replace the theoretical levels with the experimental one beyond the excitation energy for which experimental compilations have no definite spin and/or parity.

The B(GT) strength distribution for the electron capture of $^{24}$Mg $\rightarrow$ $^{24}$Na is shown in Fig. 1. Quenching of the GT strength is taken into account and a standard quenching factor of 0.77 is used [17, and reference therein]. We calculated the GT strength for 136 excited states of $^{24}$Mg up to excitation energies in the vicinity of 40 MeV in daughter $^{24}$Na. For each excited state of parent $^{24}$Mg we considered 100 states in daughter $^{24}$Na. A large model space may assist in reproducing low-lying spectrum and higher excitations [18, 19]. We employed a model space of 7ħω in our calculations. Also shown in Fig. 1 is the GT strength distribution of the shell model calculations [5]. The F$^2$N [2] and the experimental strong GT peak [20] are shown by dark circle and star, respectively, in the upper panel of Fig. 1. The GT strength at E = 0.472 MeV in the present study is equal in magnitude to the predicted shell model and experimental strength. We observed a strong peak at E = 0.97MeV. Because of lack of measurement of ground state $\beta^-$ transition, a high resolution data for transition to the mirror nucleus $^{24}$Al are available from the $(p,n)$ reaction [20]. It is well known that for the reaction on the self-conjugate nucleus, like $^{24}$Mg, the GT strength is expected to be same in both isospin direction and the symmetry equation $B(GT^-)=B(GT^+)$ can be used for comparison. We observed that this strong peak at E = 0.97 MeV is in good agreement with measured peak of $(p,n)$ reaction at E =

1.07 MeV. While in $(d, {}^2He)$ reaction, this peak is observed at E = 1.35 MeV [21] and lies within the allowed uncertainty of 0.5 MeV in calculation of the energy of nuclear state with the pn-QRPA peak.

The experimentally extracted total B(GT) strength for the considered excitation energy region of 7 MeV, where the density of states is still low enough to analyze single peaks, is 1.36 [21]. For the ${}^{24}$Mg case, the authors [21] employed sd-model space and the universal sd residual interaction of [22 23] to calculate the wave functions. The authors in [21] argued that the full model space calculation was not possible and they had to truncate the model space. This model space truncation lead to uncertainties and the authors had some reservation in the interpretation of their data. They pointed out an overall error of 30% for the $\Delta L = 0$ cross section extracted. We extracted a total B(GT) strength of 2.65 for transitions from ${}^{24}$Mg for the same excitation energy region of 7 MeV. This value is quite close to the reported shell model value of 2.1 of Brown and Wildenthal [22] and Wildenthal [23] (Takahara et al. [5] reported a value of 1.30). The Ikeda sum rule for ${}^{24}$Mg is satisfied in our calculations.

The comparison of the electron capture rates calculated by pn-QRPA (this work) and shell model is shown in Fig. 2. In low density and temperature region our rates are in good agreement with shell model rates. As the temperature proceeds toward the supernova epoch our rates are again in good agreement with those of shell model in the low density region. It is the domain of high density where our rates differ significantly with those of shell model. When the inexorable gravity shifts the core to a density of the order of $10^{11}$ g-cm$^{-3}$, our calculated rates are enhanced than the shell model rate by as much as a factor of around four (see Table I). In high density region, the Fermi energies of electrons are high enough and the electron capture rates are sensitive to the total GT strength rather than its distribution details [24]. Our total GT strength is more in comparison to shell model and results in the enhancement of our rates in high density region. We also note here that we did not employ the so-called Brink's hypothesis in our calculations as usually employed by large scale shell model calculations.

The comparison of our calculated electron capture rates and the $F^2N$ rates is shown in Fig. 3. In the low density and temperature regions, our rates are enhanced compared to the $F^2N$ rates. Two types of transitions contribute to the $F^2N$ strength function: discrete transitions to low-lying states and collective GT resonance. For the discrete transitions they compiled experimental data on the level information (excitation energy, spin and parity) and $ft$ values of β-decay and incorporated all these information in their calculations. They assumed a fixed value of $\log ft = 5.0$ for the transitions whose $ft$ values were not known. This assumption is not a good one particularly in the low density region where the Fermi energy of the electron is less than the threshold energy of the collective resonance. These assumed $ft$-values suppressed the $F^2N$ rates (in comparison to shell model as well as our rates) in the low density and temperature regions of the stellar core. At higher densities approaching $10^{11}$ g-cm$^{-3}$ the QRPA rates are in reasonable agreement with the $F^2N$ rates, albeit a bit higher. We attribute this enhancement of QRPA rates in the high density region to the total GT strength rather than its details as mentioned earlier. What could be the possible implication of our calculated electron capture rates on $^{24}$Mg in core collapse simulations and other related astrophysical processes? The O+Ne+Mg cores are gravitationally less bound than more massive progenitor stars and can release more energy due to the nuclear burning. The progenitor stars in the mass range of $8-10 M_{\odot}$ with an O+Ne+Mg core are considered as possible sites for a low entropy r-process [25, 26, 27] based on the condition that they may explode by the prompt bounce-shock mechanism. Owing to a reasonable large number of recent observations and accumulation of data regarding Type II-P supernovae, the debate on the fate of O+Ne+Mg cores was rejuvenated. Whereas Gutiérrez et al. [28] argued that the abundance of $^{24}$Mg was reduced severely in updated evolutionary calculations; the procedure adopted was not fully consistent (they kept the ratio of oxygen to neon constant while parameterizing the abundance of $^{24}$Mg). Much recently Kituaura et al. [29] presented simulation results of O+Ne+Mg cores with an improved neutrino transport treatment and found no prompt explosions, but instead a delayed explosion. Kitaura et al. [29] used the electron capture rates of Takahara et al. [5] in the non-nuclear statistical equilibrium regime. The spherically core collapse simulations [29] still do not explode partly because of the reduced electron capture, slowing the collapse and resulting in a large shock radius. The collapse simulators should take note of our enhanced microscopic calculation of electron capture rates at presupernova temperatures and high densities. This might point toward still lower value of $Y_e$ and lower entropy in the stellar core. We are in the process of finding the affects of inclusion of our rates in stellar evolution codes and hope to report soon.

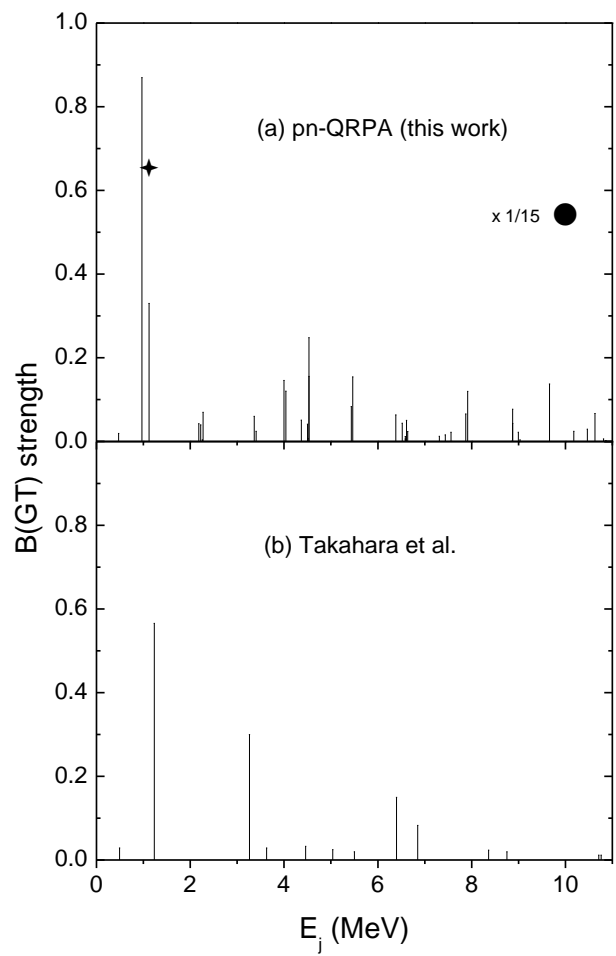

FIG. 1. Gamow-Teller strength distribution for the ground state of $^{24}$Mg. For comparison the calculated GT strength by shell model [5] is shown in lower panel. The experimental and the $F^2N$'s strong GT peak are shown by star and dark circle (adopted from [5]), respectively, in the upper panel. The energy scale refers to the excitation energy in the daughter $^{24}$Na.

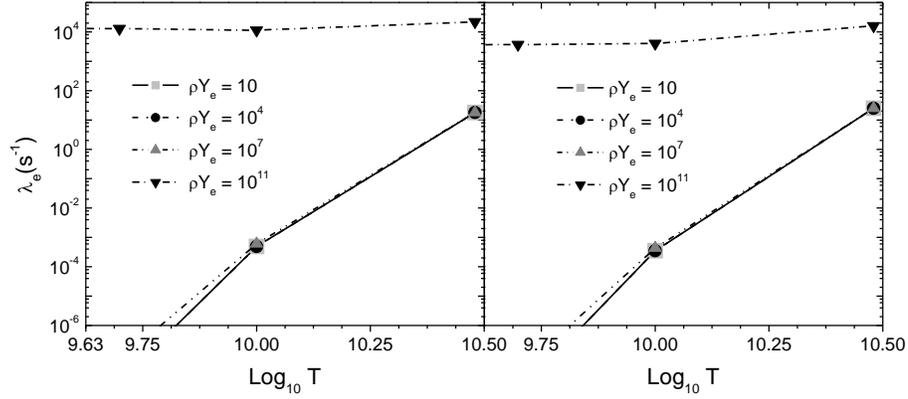

FIG. 2. Electron capture rates for $^{24}$Mg $\rightarrow$ $^{24}$Na as a function of temperature for different selected densities (left panel). The right panel shows the rates of shell model [5] for the relevant temperatures and densities. For units see text.

TABLE I. The comparison of QRPA rates and shell model (SM) rates at high temperatures and density.

| $Log_{10}T$ (K) | Density (g-cm$^{-3}$) | QRPA / SM |
|---|---|---|
| 7.0 | $10^{11}$ | 3.7 |
| 8.0 | $10^{11}$ | 3.7 |
| 9.0 | $10^{11}$ | 3.7 |
| 10.0 | $10^{11}$ | 2.8 |
| 10.5 | $10^{11}$ | 1.4 |

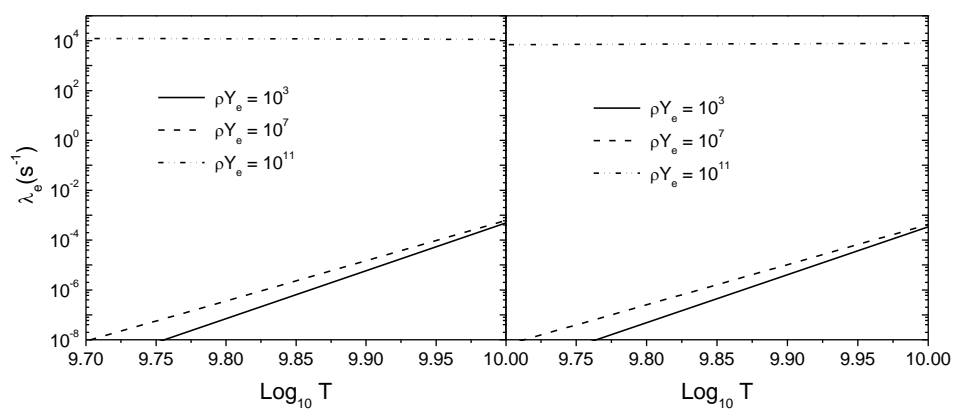

FIG. 3. Electron capture rates for $^{24}Mg \rightarrow ^{24}Na$ as a function of temperature for different selected densities (left panel). The right panel shows the rates of F$^2$N [2] for the relevant temperatures and densities. For units see text.